\documentclass[usenatbib,uAseAMS,a4paper,fleqn]{mnras}
\usepackage{amssymb}
\usepackage{newtxtext,newtxmath}
\usepackage[T1]{fontenc}
\usepackage{ae,aecompl}
\usepackage{times}
\usepackage{microtype}
\usepackage{verbatim}
\usepackage{graphicx}
\usepackage{tablefootnote}
\date{Accepted XXX. Received YYY; in original form ZZZ}
\pubyear{2021}

\usepackage[multidot]{grffile}

\begin{document}

\label{firstpage}
\pagerange{\pageref{firstpage}--\pageref{lastpage}}

\title[Dynamical evolution of HR~5183]
  {Dynamical orbital evolution scenarios of the wide-orbit eccentric planet
    HR~5183b}

\author[Mustill et al.]
       {Alexander J. Mustill$^1$\thanks{E-mail: alex@astro.lu.se},
         Melvyn B. Davies$^{1,2}$,
         Sarah Blunt$^{3,4}$,
         Andrew Howard$^3$\\
         $^1$Lund Observatory, Department of Astronomy \& Theoretical Physics,
         Lund University, Box 43, SE-221 00 Lund, Sweden\\
         $^2$Centre for Mathematical Sciences, Lund University, Box 118, SE-221 00 Lund, Sweden\\
         $^3$Department of Astronomy, California Institute of Technology, Pasadena, CA, USA\\
         $^4$NSF Graduate Research Fellow
       }

\maketitle

\begin{abstract}
  The recently-discovered giant exoplanet HR5183b exists on a wide, 
  highly-eccentric orbit ($a=18$\,au, $e=0.84$). Its host star possesses a 
  common proper-motion companion which is likely on a bound orbit. 
  In this paper, we explore scenarios for the excitation of the eccentricity 
  of the planet in binary systems such as this, 
  considering planet--planet scattering,
  Lidov--Kozai cycles from the binary acting on a single-planet
  system, or Lidov--Kozai cycles acting on a two-planet system that 
  also undergoes scattering. 
  Planet--planet scattering, in the absence of 
  a binary companion, has a $2.8-7.2\%$ probability of
  pumping eccentricities to the observed values 
  in our simulations, depending on the relative masses of the two planets.
  Lidov--Kozai cycles from the binary acting on an initially 
  circular orbit can excite eccentricities to the observed value, 
  but require very specific orbital configurations for the binary 
  and overall there is a low probability of catching 
  the orbit at the high observed high eccentricity ($0.6\%$). 
  The best case is 
  provided by planet--planet scattering in the presence of a binary companion: 
  here, the scattering provides the surviving planet with an initial eccentricity 
  boost that is subsequently further increased by Kozai cycles from the binary.
  We find a success rate of $14.5\%$ for currently observing $e\ge0.84$ in this set-up. 
  The single-planet plus binary and two-planet plus binary cases are 
  potentially distinguishable if the mutual inclination of the binary 
  and the planet can be measured, as the latter permits a broader range 
  of mutual inclinations. The combination of scattering and Lidov--Kozai forcing 
  may also be at work in other wide-orbit eccentric giant planets, which have 
  a high rate of stellar binary companions. 
\end{abstract}


\section{Introduction}


The G0 star HR~5183 ($M_\star=1.07\pm0.04\mathrm{\,M}_\odot$), 
has been found to host a super-Jovian 
planet ($M\sin I=3.23^{+0.58}_{-0.55}\mathrm{\,M_J}$) 
on a highly-eccentric orbit \citep{Blunt+19}.
While the orbit of the planet is wide and its semimajor axis
poorly constrained ($a=18^{+6}_{-4}$\,au), good coverage 
of the planet's periastron passage both permitted its discovery and 
provided a good determination of the orbital eccentricity 
($e=0.84\pm0.04$).

The high eccentricity of HR~5183b suggests that significant 
changes to the planet's orbit have happened in the past, as 
planets form on near-circular orbits\footnote{Gravitational instability 
in a massive disc can sometimes create fragments 
with high eccentricity, but these often result from 
interactions between multiple fragments in the same system -- essentially 
planet--planet scattering at an early stage 
\citep{Hall+17,Forgan+18}.}. Planet--planet 
scattering has been shown to be able to produce high eccentricities
\citep{RasioFord96,WeidenschillingMarzari96,Chatterjee+08,JuricTremaine08}.
The eccentricity distribution of giant planets 
implies that around 80\% of them formed in unstable 
multi-planet systems \citep{JuricTremaine08,Raymond+11}. 
Planet--planet scattering can excite eccentricity even to near-unity 
values \citep{Carrera+19}; and the more massive the planet ejected, 
the larger the eccentricity of the survivor 
\citep[e.g.,][figure~1]{Kokaia+20}. Thus, the high eccentricity 
of HR~5183b would suggest that it formed in a system with at 
least one other super-Jovian planet, of which it is the only survivor.

\begin{figure}
  \includegraphics[width=0.5\textwidth]{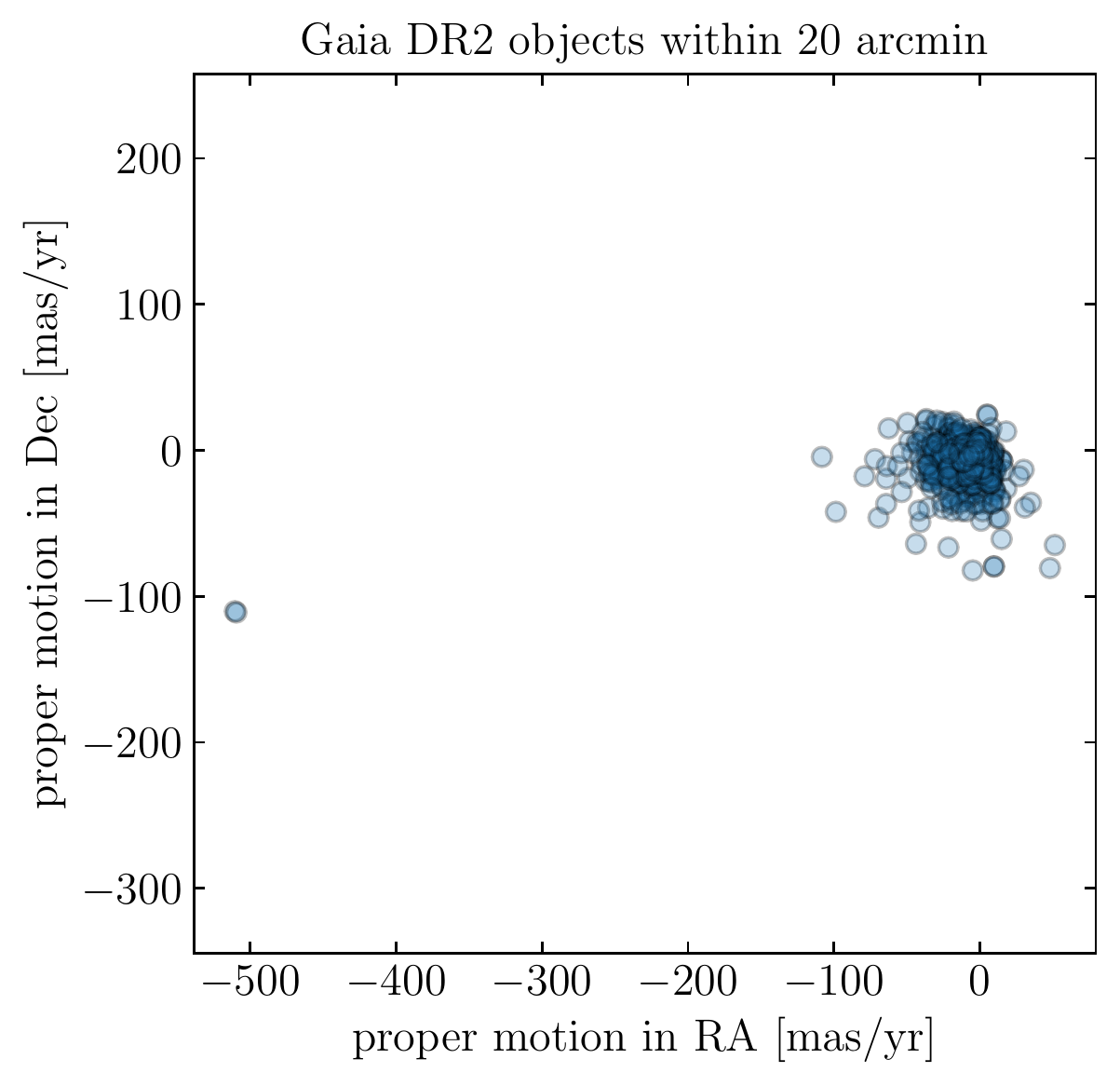}
  \caption{Proper motions of all objects in \emph{Gaia} DR2 
    within 20 arcmin of HR~5183. HR~5183 and HIP~67291 are 
    the two barely-distinguishable points at around $(-500,-100)$. 
    Their remoteness from the main cloud of field objects 
    means that the probablility that the pair is a 
    chance alignment of unbound objects is vanishingly small.}
  \label{fig:pm}
\end{figure}

However, HR~5183 possesses a common-proper motion 
companion \citep{Allen+00}. 
Figure~\ref{fig:pm} shows the proper motions of 
all objects within 20 arcmin of HR~5183 from \emph{Gaia} DR2
\citep{Gaia+16,Gaia+18}, 
underlining the extremely low likelihood that this is a chance 
alignment. Furthermore, the parallax and radial velocity of 
the two stars are almost identical \citep{Blunt+19}. This star,
of mass $M=0.67\pm0.05\mathrm{\,M}_\odot$, 
has a projected separation of 15\,000\,au. 
The high-quality astrometric data from \emph{Gaia} 
allowed \cite{Blunt+19} to constrain the orbital 
elements of the binary: 
the distribution of semi-major axis and eccentricity (for bound orbits) 
is shown in the left-hand panel of Figure~\ref{fig:binaries}. 
Details of the HR~5183 system are given in Table~\ref{tab:HR5183}. 
While too distant to affect the protoplanetary disc 
where planet formation takes place \citep{Harris+12},
the presence of a binary companion would complicate the 
subsequent planet--planet scattering described above: it could trigger 
an instability in an otherwise stable system, and/or 
it could affect the orbit of a planet that survives 
the scattering process.

\begin{table*}
  \centering
  \caption{Relevant parameters for the HR~5183 system from
    \protect\cite{Blunt+19}, together with values 
    adopted for the simulations.}
  \label{tab:HR5183}
  \begin{tabular}{lcc}
    Parameter & Blunt et al.\ (2019) value & Adopted value \\
    \hline
    Primary mass $M_\star$ & $1.07^{+0.04}_{-0.04}\mathrm{\,M}_\odot$ & $1.07\mathrm{\,M}_\odot$\\
    System age $t_\mathrm{age}$ & $7.7^{+1.4}_{-1.2}$\,Gyr & $8$\,Gyr\\
    \hline
    Secondary mass $M_\mathrm{B}$ & $0.67^{+0.05}_{-0.05}\mathrm{\,M}_\odot$ & $0.67\mathrm{\,M}_\odot$\\
    Binary semimajor axis$^1$ $a_\mathrm{B}$ & see Figure~\ref{fig:binaries} & drawn from\\
    Binary eccentricity$^1$ $e_\mathrm{B}$ & see Figure~\ref{fig:binaries} & joint posterior\\
    \hline
    Planet minimum mass $M_\mathrm{b}\sin I$ & $3.23^{+0.15}_{-0.14}\mathrm{\,M_J}$ & $M_\mathrm{b}=3.23\mathrm{\,M_J}$\\
    Planet semimajor axis $a_\mathrm{b}$ & $18^{+6}_{-4}$\,au & $18$\,au \\
    Planet eccentricity $e_\mathrm{b}$& $0.84\pm0.04$ & $0.84$\\
    \hline
    \multicolumn{2}{l}{$^1$ restricting to bound orbits $<200\,000$\,au.}
  \end{tabular}
\end{table*}

Wide binary companions can excite an exoplanet's eccentricity 
from an initial low value 
through Lidov--Kozai cycles, if the binary's orbital 
inclination is sufficiently high \citep{Kozai62,Lidov62}. 
However, the higher the eccentricity one wishes to 
attain, the higher the initial 
inclination required; furthermore, planets only spend a short 
time at the peak of the eccentricity cycle. As we show below, 
these factors mean that it is unlikely that we are observing 
the planet at its present high eccentricity if the dynamics 
responsible were purely Lidov--Kozai excitation from 
a near-circular initial orbit.

In this Paper, we therefore consider the two effects, of planet--planet 
scattering and Lidov--Kozai cycles from a binary companion,
operating in concert. 
In otherwise stable multi-planet systems, planet--planet interactions 
can suppress the Lidov--Kozai cycles
\citep{Innanen+97,Malmberg+07,Kaib+11,BoueFabrycky14,Mustill+17,
  PuLai18,Denham+19};
alternatively, the binary companion can destabilise 
the planetary system by inducing a large eccentricity on the outermost planet                               
\citep{Malmberg+07,Mustill+17}. The binary destabilises 
the system if the timescale for Lidov--Kozai cycles is 
shorter than that for secular planet--planet interactions. 
On the other hand, a multi-planet system may be closely spaced 
enough to undergo an internal instability, without external 
triggering \citep{Gladman93,Chambers+96,Quillen11,Petit+20}. In this case, 
the binary companion can subsequently further excite the eccentricity
of a planet that survives the instability \citep{Kaib+13}.
We show in this paper
that this significantly increases the probability of observing a planet 
at a high eccentricity, above either planet--planet scattering 
alone or Lidov--Kozai cycles alone.


This paper is organised as follows: in Section~\ref{sec:sims} 
we describe the set-up of our three sets of $N$-body simulations; 
in Section~\ref{sec:results} 
we present the outcomes of the simulations; 
and in Section~\ref{sec:discuss} 
we discuss the results.
We conclude in Section~\ref{sec:conclude}.

\section{Simulation set-up}

\label{sec:sims}

\begin{figure*}
  \includegraphics[width=0.48\textwidth]{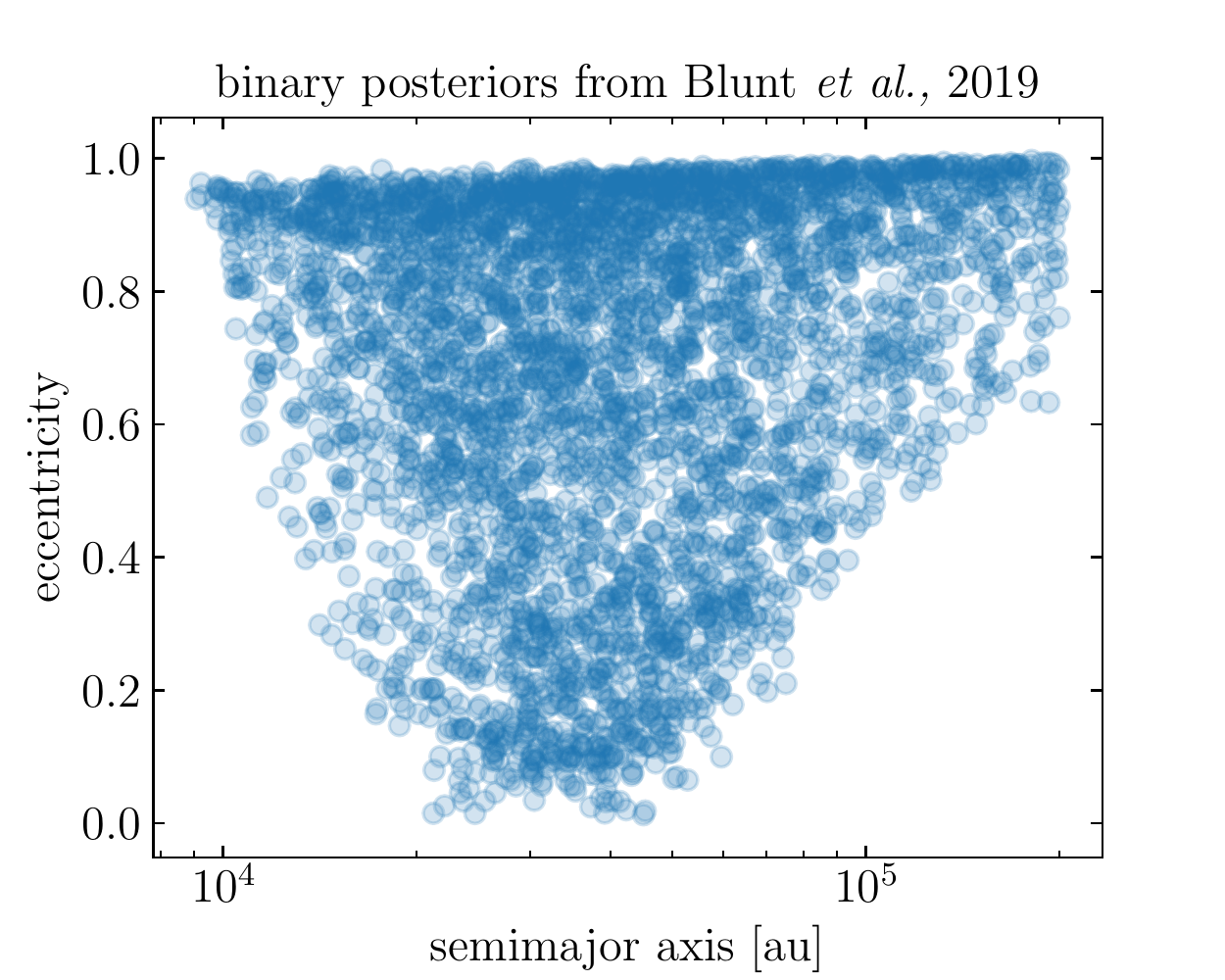}
  \includegraphics[width=0.48\textwidth]{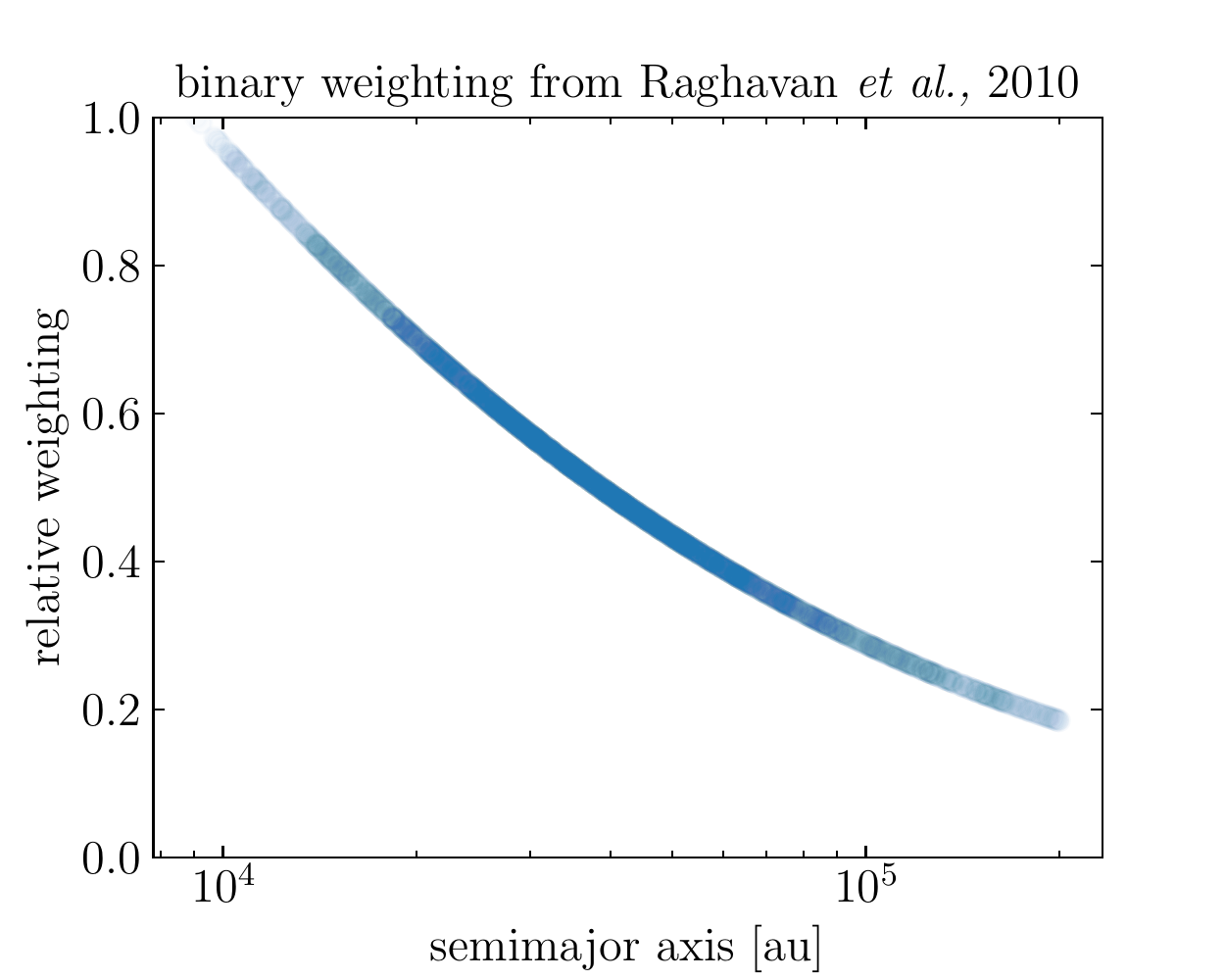}
  \caption{\textbf{Left:} Binary orbital parameters from \protect\cite{Blunt+19}. 
  3\,936 draws from the posterior distribution are shown, 
  restricted to bound orbits with semimajor axis less than 200\,000\,au. 
  These orbits are then sampled to draw the binary companions 
  for the $N$-body simulations. \textbf{Right: }Weighting applied 
  when sampling from the binary orbits from \protect\cite{Blunt+19} in 
  the \textsc{weighted} runs; the weighting is from the observed lognormal 
  binary period distribution of \protect\cite{Raghavan+10}.}
  \label{fig:binaries}
\end{figure*}

We conduct several sets of $N$-body simulations to explore the origin of 
the high eccentricity of HR~5183b. The simulations were all 
run with \textsc{Mercury} \citep{Chambers99} using the accurate 
RADAU integrator \citep{Everhart85}, with an error tolerance of $10^{-11}$. 
We adopted the values 
$M_\star=1.07\mathrm{\,M}_\odot$ and $R_\star=1.53\mathrm{\,R}_\odot$ 
from \cite{Blunt+19}. Collisions between planets, or between a planet 
and a star, are treated as perfect inelastic mergers.

We run three sets of simulations: pure scattering, with 
the primary star and two planets; pure Kozai, with 
the primary star, one planet, and the binary companion; 
and combined scattering and Kozai, with the primary star, 
two planets and the binary companion:
\begin{enumerate}
\item \emph{Pure scattering:} Here we consider the simplest 
  case of planet--planet scattering, and set up systems each 
  with two planets.
  One set of runs has equal planet masses 
  $M_1=M_2=3.23\mathrm{\,M_J}$, and three sets have 
  planet--planet mass ratios $\mu=1.2$, $1.5$ and 
  $2.0$, where the smaller planet has mass 
  $M_1=3.23\mathrm{\,M_J}$. 
  The inner planet is placed at $a_1=36\mathrm{\,au}$; this means 
  that its semimajor axis shrinks to roughly the observed 
  $18$\,au after ejection of the comparable-mass outer planet. The 
  exact final semimajor axis is not significant, because at 
  these separations with very rare physical collisions the 
  dynamics is essentially scale-invariant.
  The outer planet itself is placed at 
  $a_2\in a_1\times[1,1+3.6((M_1+M_2)/M_\star)^{1/3}]$, thus 
  extending beyond the Hill stability boundary \citep{Gladman93}
  in order to include systems which are Hill stable but 
  Lagrange unstable \citep{VerasMustill13}\footnote{I.e., 
  the planets are protected from orbit-crossing, but one 
  can still escape to infinity through successive weak 
  encounters.}. 
  In the unequal-mass 
  cases the more massive planet is exterior to encourage its ejection 
  and help the surviving planet attain a higher eccentricity. 
  Planets are started on circular orbits with a small 
  mutual inclination of up to $1^\circ$.
  Systems are integrated for 1\,Gyr (most instabilities 
  occur early, and very few at such late times) or until the loss of 
  a planet by collision or ejection from the system. Thus these systems 
  extend to several 10s of au; while rare, direct imaging 
  surveys show that systems of multiple giant planets at these orbital radii 
  and beyond do indeed exist \citep[e.g.,][]{Marois+08,Bohn+20}.
\item \emph{Pure Kozai: }
  Here we place one planet at 18\,au and add a binary stellar companion 
  drawn from the posterior orbit fits of \citet[see below]{Blunt+19}. 
  The binary orbit is oriented isotropically with respect 
  to that of the planet (while we have constraints on the present binary 
  orbital inclination, we are at this stage agnostic as 
  to the initial orbital inclination of the planet). 
  The binary star's mass is $0.67\mathrm{\,M}_\odot$.
  Systems are integrated for the system age of 8\,Gyr.
\item \emph{Combined scattering and Kozai: } here we place two near-coplanar
  equal-mass planets in each system, as in the pure scattering runs, 
  and add a binary stellar companion as in the Kozai runs. 
  Systems are again integrated for the system age of 8\,Gyr.
  We note that in principle the addition of the binary
    companion could act to destabilise the planetary
    system as Lidov--Kozai cycles could be induced on the
    outermost planet, making scattering more likely
    \citep{Innanen+97,Malmberg+07,PuLai18,Denham+19}. However,
    the binary companions in our systems are too widely spaced:
    the timescale for Lidov--Kozai interactions is typically
    $\gtrsim1$\,Gyr, and these are therefore suppressed by
    planet--planet interactions which have timescales
    $\lesssim1$\,Myr. These combined systems therefore
    evolve first under planet--planet scattering, and then under
    Lidov--Kozai perturbations from the companion.
\end{enumerate}

For the runs with binaries, we run both \textsc{Unweighted} 
and \textsc{Weighted} sets. In the \textsc{Unweighted} runs 
we draw the binary semimajor axis and eccentricity from the 
posterior distributions of \cite{Blunt+19}, restricting 
ourselves to bound orbits with semimajor axes $<200\,000$\,au 
($\sim1$\,pc, taken as a crude upper limit of where a binary 
would avoid being broken up in the Galactic field); 
draws from this distribution are shown in the left-hand 
panel of Figure~\ref{fig:binaries}. These posteriors derive 
from the reported formal errors in \emph{Gaia}~DR2, 
not accounting for possible systematics. 
These posterior distributions also had a uniform prior on the binary 
semimajor axis distribution. However, the period distribution of 
the binary population for Solar-type stars 
appears to be lognormal with a mean of $\log (P/\mathrm{d})=5.03$
and a standard deviation of $\sigma_{\log P/\mathrm{d}}=2.28$ 
\citep{Raghavan+10}; \cite{TokovininLepine12} 
show that this distribution extends to 
the very wide binaries beyond $\sim10\,000$\,au that we consider here. 
We therefore construct \textsc{Weighted} sets of simulations where 
the posterior from \cite{Blunt+19} is weighted by the 
binary semimajor axis with the lognormal weighting function 
shown in the right-hand panel of Figure~\ref{fig:binaries}.

\section{Results}

\label{sec:results}

\begin{figure*}
  \includegraphics[width=0.33\textwidth]{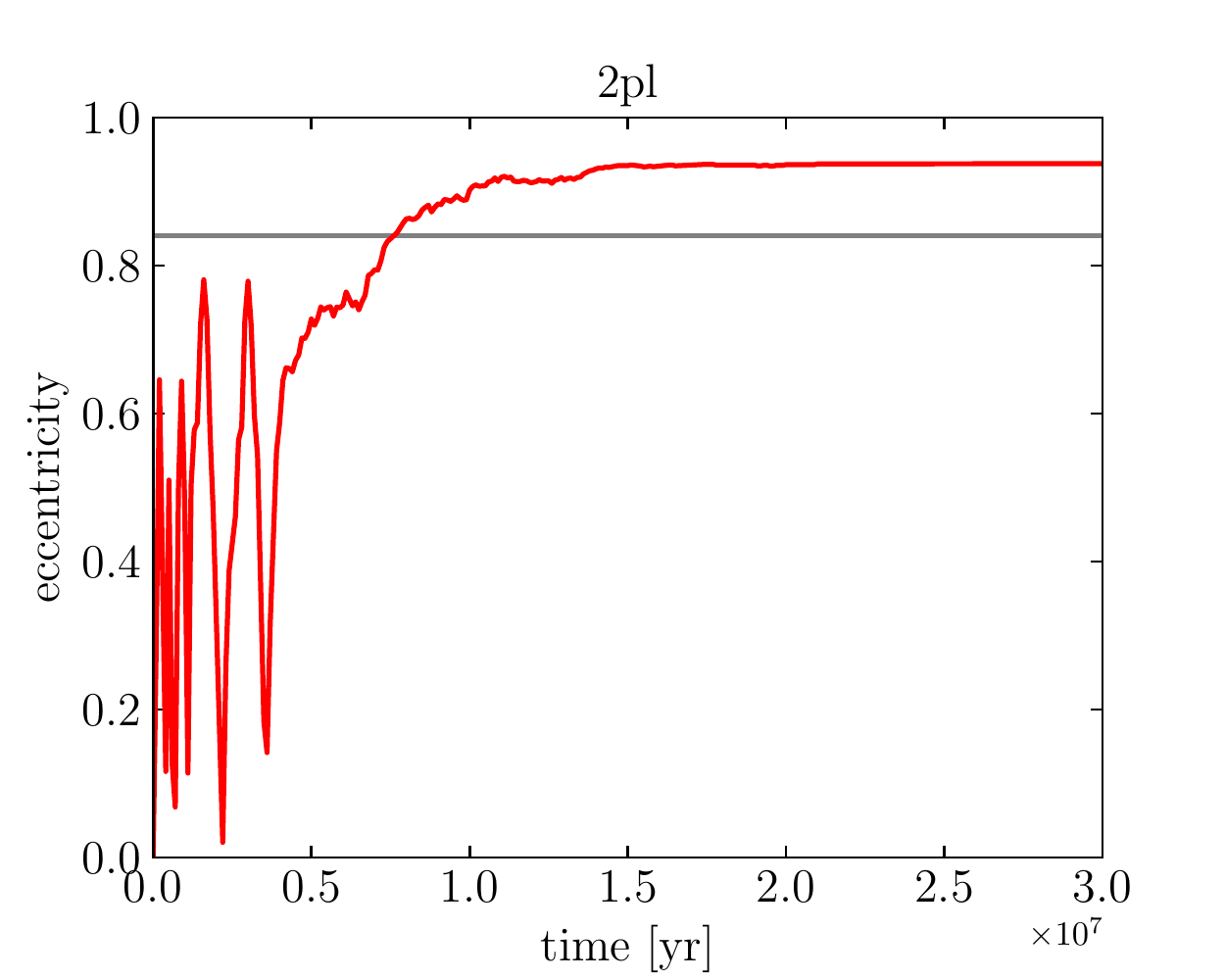}
  \includegraphics[width=0.33\textwidth]{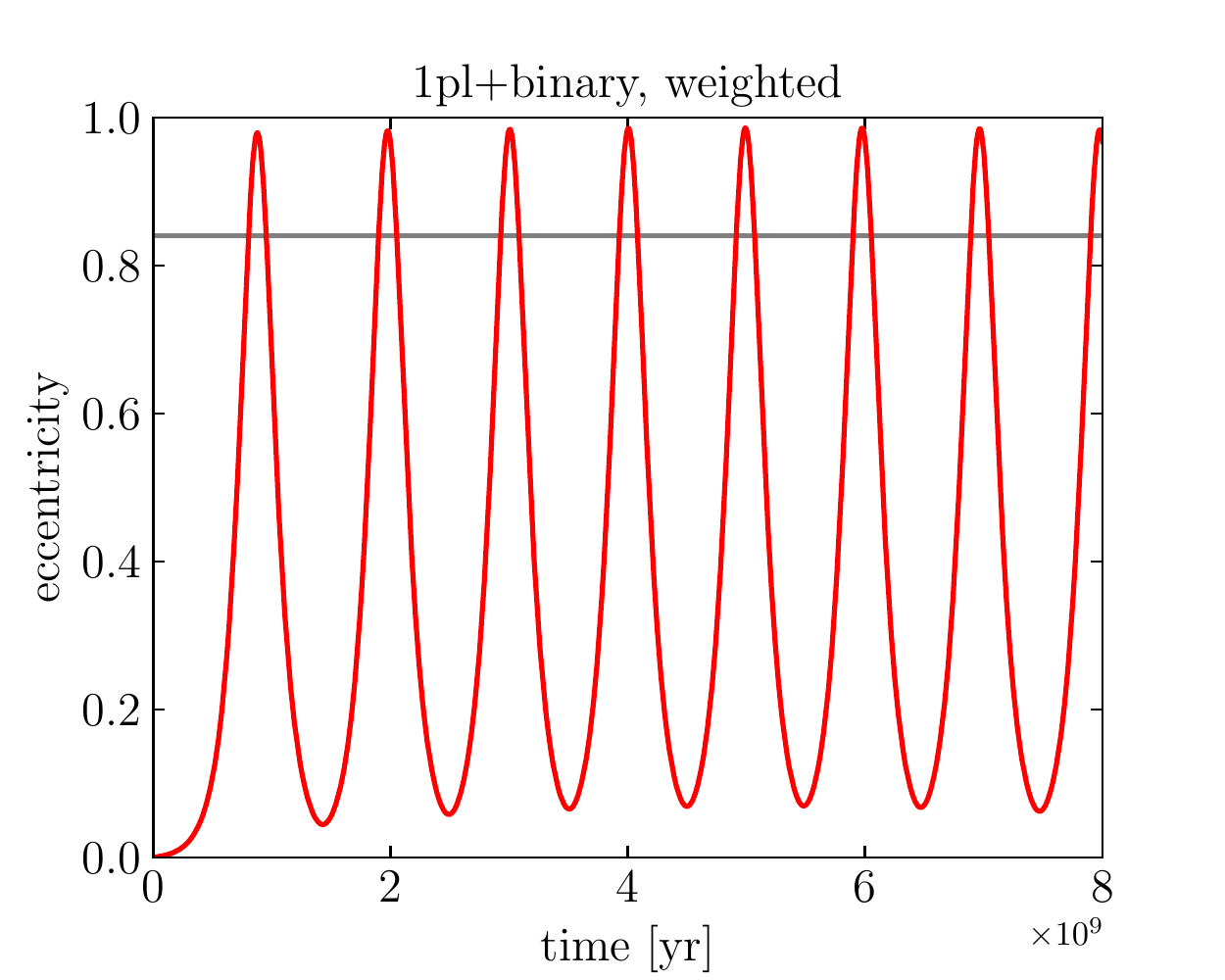}
  \includegraphics[width=0.33\textwidth]{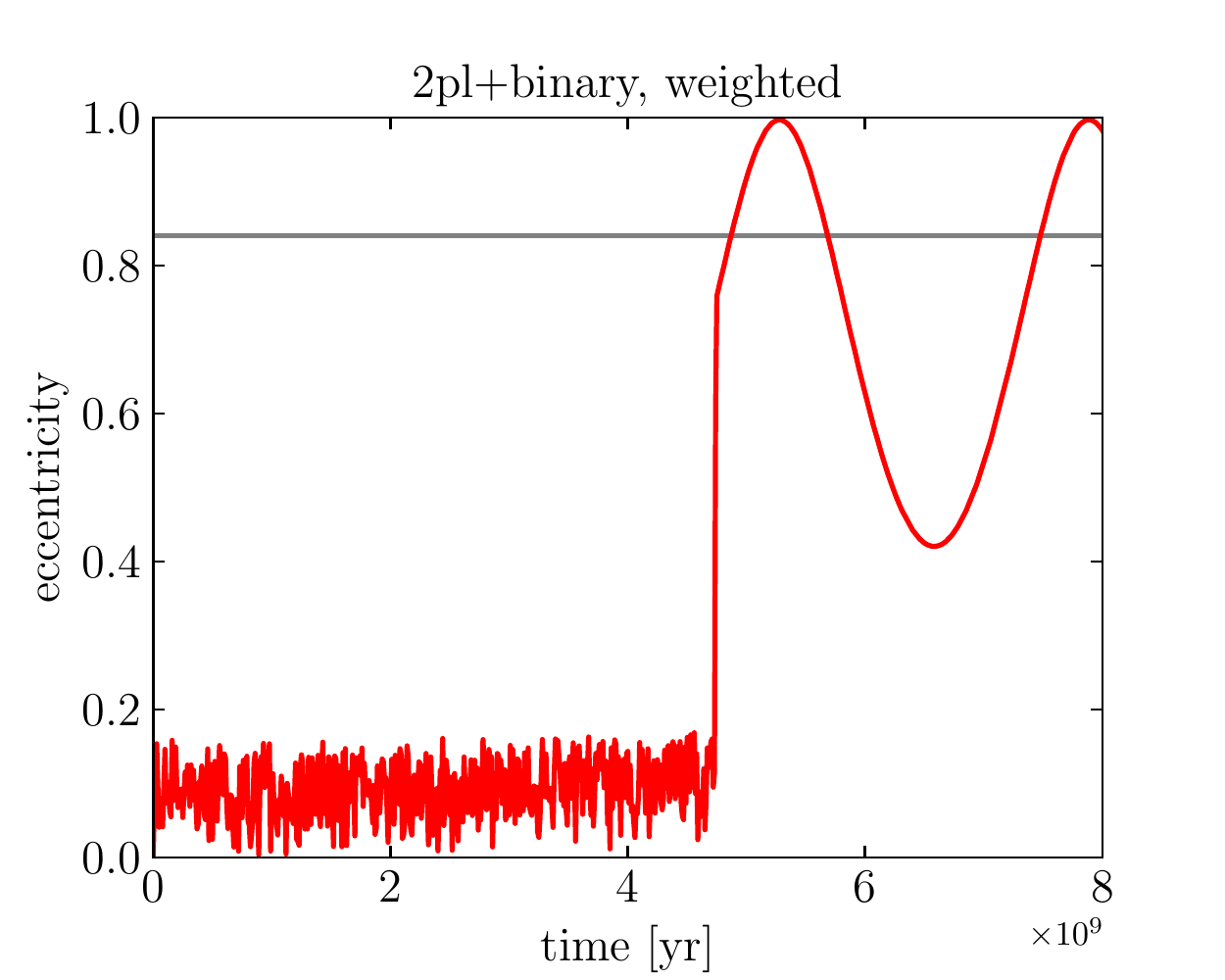}
  \caption{Examples of eccentricity evolution for the
    surviving planet in three runs where $e>0.84$ at
    the end of the simulation. Left: Two planets with
    no binary companion. After an initial phase of scattering the second
    planet is ejected at 25\,Myr, freezing in the surviving planet's
    eccentricity at a high value. Centre: Single planet with binary
    companion. The planet's eccentricity cycles up and down
    under the influence of Lidov--Kozai perturbations. We are fortunate
    enough to catch the system in a high-eccentricity phase at the
    end of the simulation. Right: Two planets with a binary companion.
    After scattering ust before 5\,Gyr, the surviving planet
    undergoes Lidov--Kozai oscillations imposed by the binary.
    While a favourable phase is still required to observe
    the planet at a high eccentricity at the epoch of observation,
    note that the ecentricity is greater than that observed over a much
    greater fraction of the cycle than in the case of the single planet
    plus companion. In each panel, we mark the target eccentricity of
    $0.84$ with a horizontal grey line.}
  \label{fig:egs}
\end{figure*}

First, we define for all simulations our success 
criterion: the existence of a planet at an eccentricity 
greater than or equal to that observed at the end of the 
simulation: $e_\mathrm{final}\ge0.84$. When calculating 
fractions of successful runs in the simulations containing two planets,
we only count systems which actually underwent 
an instability. This avoids our having to test different
distributions of initial orbital spacing, by which we could 
make the fraction of unstable systems arbitrarily large or small.
We choose not to impose a semi-major axis cut on the surviving
  planets for two reasons. Firstly, at these distances from the star, collisions between
  bodies are extremely unlikely, and so a system undergoing scattering
  can be rescaled to match a given semimajor axis. Second, we lack a predictive
  theory of planet formation, and observations are not as yet very constraining
  about the planet population at tens of au, so it is hard to know what
  initial conditions in semimajor axis should be. However, it is much more
  certain that planets form on low-eccentricity orbits, and hence the
  current eccentricity provides the strongest useful constraint on
  the system's dynamical history.

\subsection{Pure scattering}

\begin{figure}
  \includegraphics[width=0.5\textwidth]{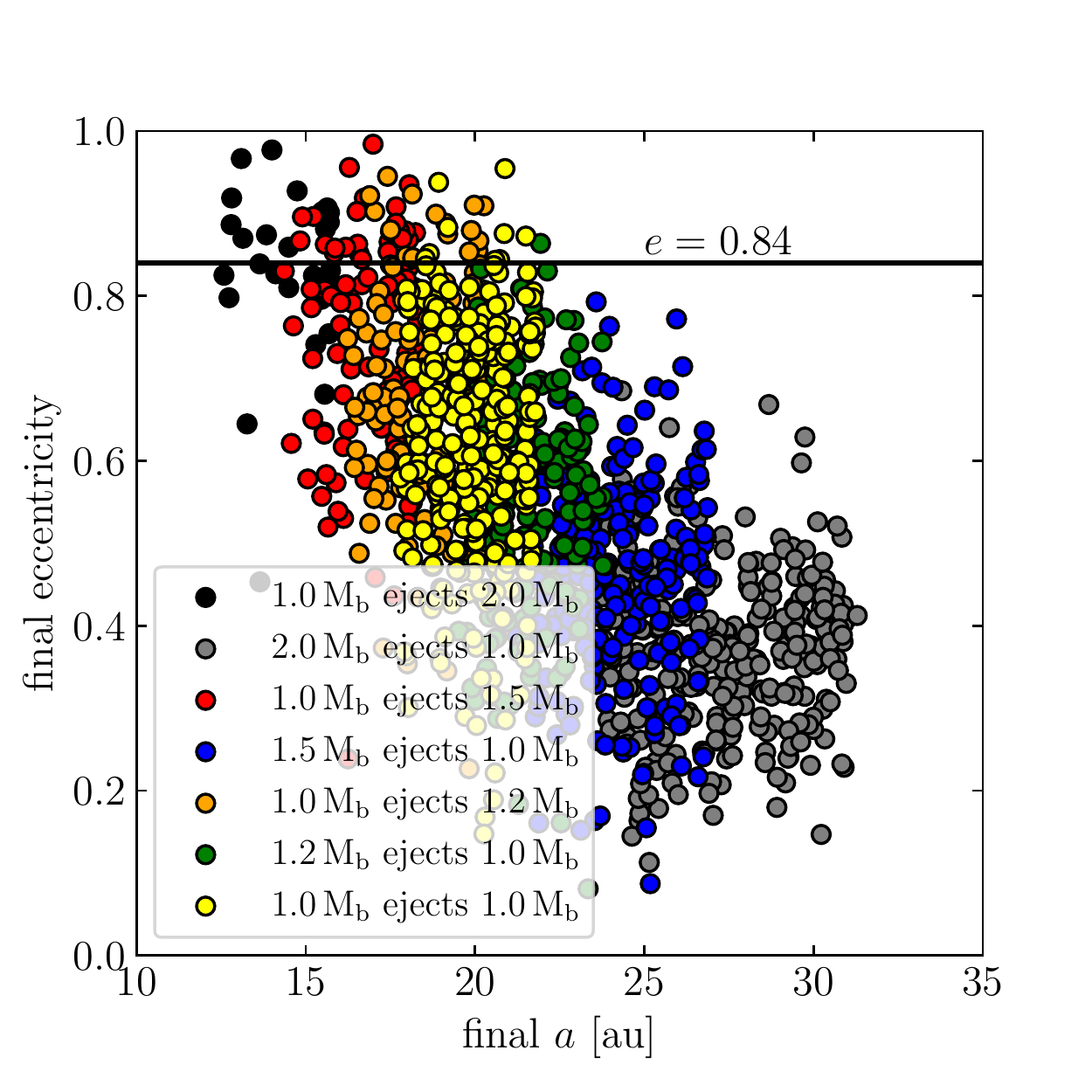}
  \caption{Orbital elements of the surviving planet 
    after instability in unstable two-planet systems with no binary 
    companion. Different colours indicate different ratios of 
    ejected to retained planet mass. The black horizontal line
  marks the target eccentricity, $e=0.84$. }
  \label{fig:2pl}
\end{figure}

\begin{table}
  \centering
  \caption{Results of the two-planet runs, with no 
    binary companion. We show the mass ratio of the 
    two planets, the number of simulations $n_\mathrm{runs}$;
    the number that were unstable $n_\mathrm{unstable}$; 
    the number where the surviving planet had an 
    eccentricity above $0.84$ at the end of the simulation 
    $n_{e\ge0.84}$; and the success fraction 
    $f_\mathrm{success} = n_{e\ge0.84}/n_\mathrm{unstable}$.}
  \label{tab:2pl-results}
  \begin{tabular}{lccccc}\\ 
    Simulation set & $m_\mathrm{out}/m_\mathrm{in}$ & $n_\mathrm{runs}$ &
    $n_\mathrm{unstable}$ & $n_{e\ge0.84}$ & $f_\mathrm{success}$\\
    \hline
    2pl-mu1.0 & 1.0 &  499$^1$  &  324  &   9  &  2.8\%  \\
    2pl-mu1.2 & 1.2 &  500  &  359  &  19  &  5.3\%\\
    2pl-mu1.5 & 1.5 &  500  &  376  &  27  &  7.2\%\\
    2pl-mu2.0 & 2.0 &  500  &  370  &  14  &  3.8\%\\
    \multicolumn{6}{p{.95\linewidth}}{$^1$ One run with bad energy conservation 
      was removed ($|\mathrm{d}E/E|>10^{-3}$). Other typical energy errors 
      were $\sim10^{-5}$ or less for this simulation set, 
      and much lower for the other sets.}
  \end{tabular}
\end{table}

Our two-planet systems are initially tightly packed,
resulting in most of them ($65-75\%$) being unstable 
(see Table~\ref{tab:2pl-results}). If a system is unstable, 
the most common outcome is the ejection of one planet, 
usually the least massive. Physical collisions between the 
planets are rare because of the small physical radius 
compared to the large orbital radius. These 
planet--planet collisions invariably result in 
moderately low eccentricities 
of below $0.2$.

We show an example of the eccentricity evolution
  of a planet that attains the target eccentricity in the left-hand
  panel of Figure~\ref{fig:egs}. The system rapidly becomes unstable
  and the second planet is ejected at $~\sim25$\,Myr. At this point,
  the surviving planet's eccentricity, which has been excited to
  above $0.9$ by the scattering, is frozen at this high value and remains
  unchanged for the rest of the system's history, as the
  origin of the sole perturbing force has been removed.

\begin{figure}
  \includegraphics[width=0.5\textwidth]{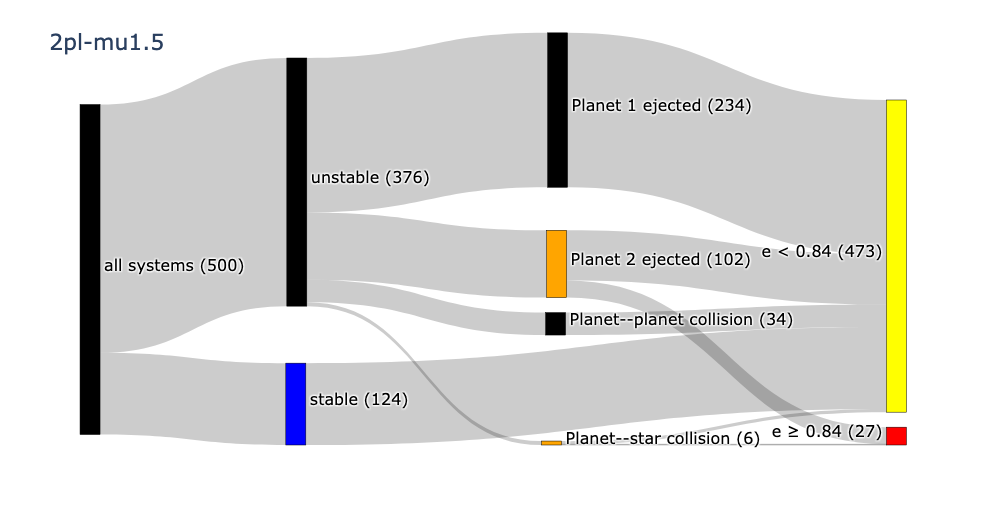}
  \includegraphics[width=0.5\textwidth]{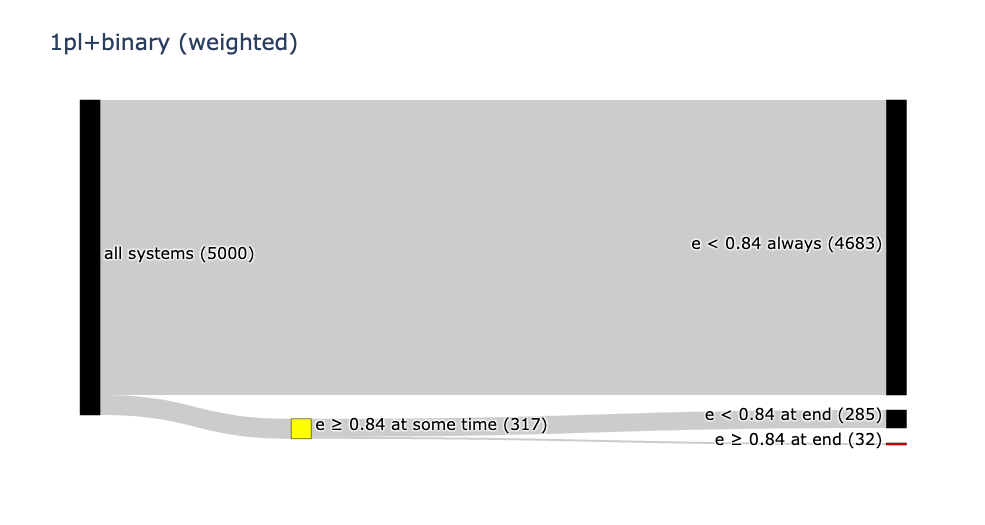}
  \includegraphics[width=0.5\textwidth]{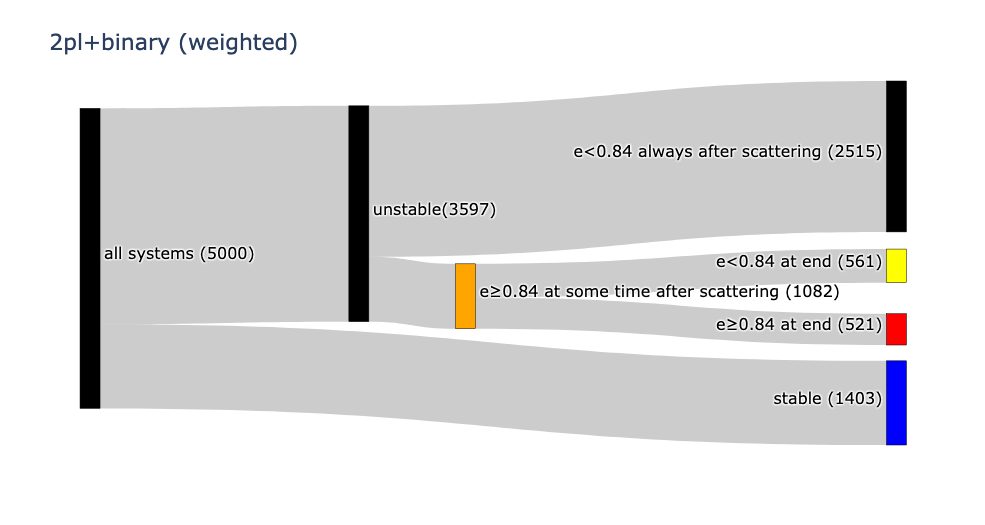}
  \caption{Simulation outcomes for our three scenarios. 
      Top: Two-planet scattering simulations 
      with no binary companion, for a planet--planet mass ratio of
      $\mu=m_\mathrm{out}/m_\mathrm{in}=1.5$
      (the most favourable case for reaching the target eccentricity).
      Middle: Simulations with a single planet
      plus a binary companion drawn from the weighted posteriors
      of \protect\cite{Blunt+19}.
      Bottom: Simulations with a two equal-mass planets
      plus a binary companion drawn from the weighted posteriors
      of \protect\cite{Blunt+19}. Note the much larger
      fraction of systems with a planet at or above the target eccentricity
      in this case (red bar).}
  \label{fig:sankey}
\end{figure}

The orbital elements of the surviving planets following 
an ejection are shown in Figure~\ref{fig:2pl}. 
A planet ejects a body of similar mass and 
initial orbital radius, so its semimajor axis roughly halves 
to conserve energy, ending at around 18\,au. 
Eccentricities are moderate to high, but 
few are at the observed value or higher, with 
success rates for $e\ge0.84$ between 3 and 7\% (Table~\ref{tab:2pl-results}). 
The eccentricity distribution is 
higher the more massive is the ejected planet,
but ejection of a more massive planet is decreasingly unlikely with 
increasing mass. These tendencies work against each other 
and the most successful mass ratio is $1.5$ 
(Table~\ref{tab:2pl-results}). A Sankey diagram 
showing how different outcomes are obtained
for this mass ratio is 
shown in Figure~\ref{fig:sankey} (top panel). 
We see that although loss of the less massive 
planet is the dominant outcome of instability, 
all but one of the highly-eccentric planets are actually 
from systems where the more massive planet was 
ejected. The exception comes from one case where the less massive planet 
collided with the star, leaving the more massive 
survivor on a very wide orbit of several hundred au. 
  
\subsection{Pure Kozai}

\begin{figure}
  \includegraphics[width=0.5\textwidth]{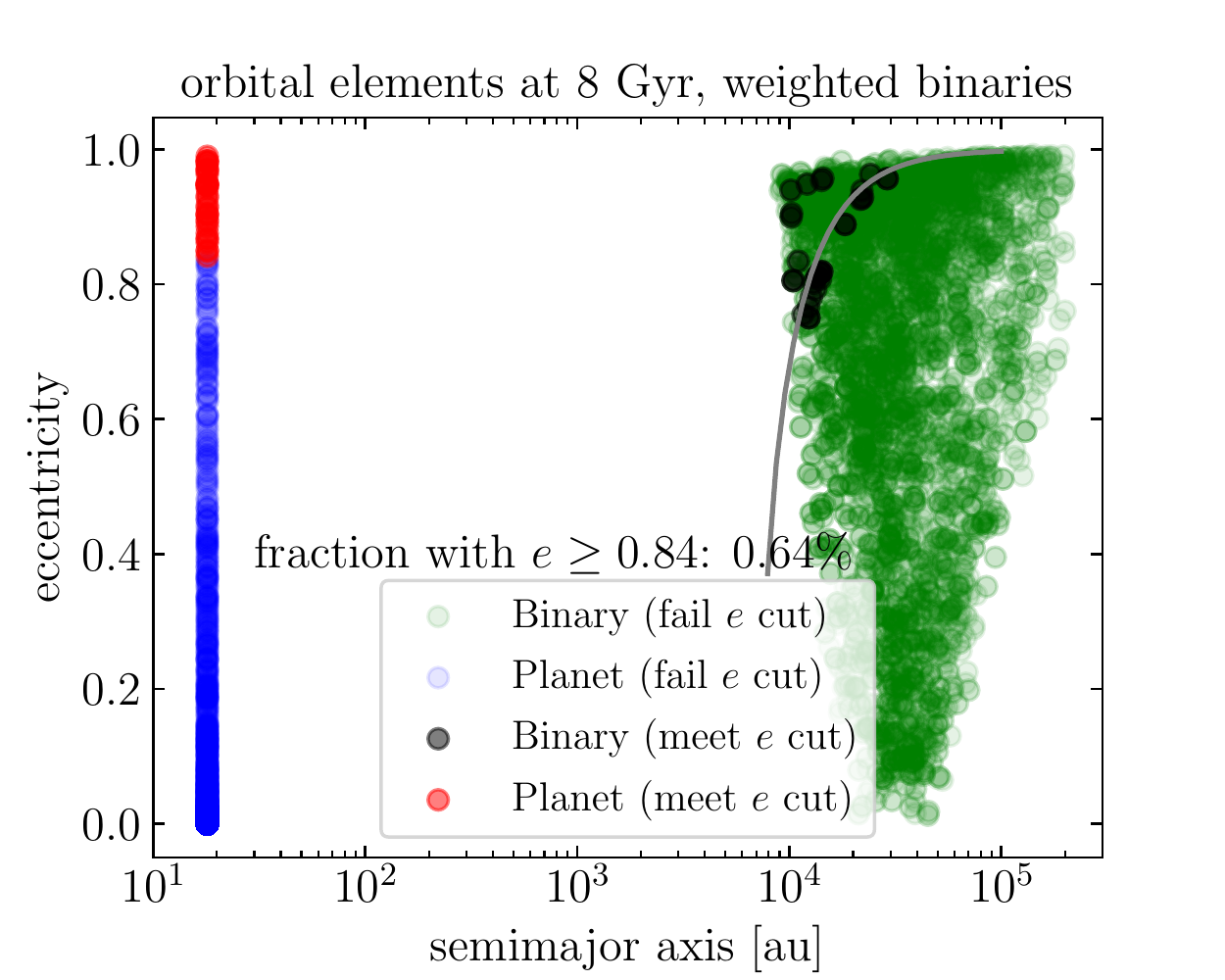}
  \caption{Orbital elements at 8\,Gyr for the 
    systems with a single planet, together with binary 
    companions drawn from the weighted posteriors of \protect\cite{Blunt+19}.
    The grey line shows where the characteristic Kozai
    timescale (Equation~\ref{eq:tKoz}) is $8$\,Gyr.
    5\,000 simulations are run.}
  \label{fig:1pl+binary}
\end{figure} 

\begin{table}
  \centering
  \caption{Outcomes for the runs with one planet 
    with a binary companion. For each simulation set 
    we show the number of runs $n_\mathrm{runs}$; 
    the number where the planet attained an 
    eccentricity above $0.84$ at any point in the 
    integration $n_{e\ge0.84}$(ever); the number 
    where the planet had an eccentricity greater than 
    $0.84$ at the end of the simulation 
    $n_{e\ge0.84}$(8 Gyr); and the success fraction 
    $f_\mathrm{success}=n_{e\ge0.84}\textrm{(8\,Gyr)}/n_\mathrm{runs}$.}
  \label{tab:1pl-kozai}
  \begin{tabular}{lcccc}
    Simulation set & $n_\mathrm{runs}$ & $n_{e\ge0.84}$ & 
      $n_{e\ge0.84}$ & $f_\mathrm{success}$\\
    & & (ever) & (8\,Gyr) & \\
      \hline
      1pl+binary-unweighted &  500 &  13 &  2 & 0.4\%\\
      1pl+binary-weighted   & 5000 & 317 & 32 & 0.6\%
  \end{tabular}
\end{table}

An example of one of our successful single-planet
  Kozai runs is shown in the centre panel of Figure~\ref{fig:egs}.
  The highly-inclined ($81^\circ$) binary can excite
  eccentricities well above $0.9$ in the planetary orbit,
  and we are fortunate in this case to ``observe'' the system
  at 8\,Gyr when it is at a high-eccentricity phase of the cycle.
  Note, however, that for most of the cycle the eccentricity
  is lower than the target value.

The presence of a sufficiently inclined binary companion 
induces oscillations in the planet's eccentricity, 
with higher eccentricities being attained when the 
binary's inclination is higher. In our simulations, 
$3-6\%$ of planets ever attain an eccentricity 
of $0.84$ or above (Table~\ref{tab:1pl-kozai}). 
As we have seen, however, and
in contrast to the case of two-planet 
scattering (where orbital evolution ceases following 
the ejection of one planet), when the eccentricity 
excitation is due to a binary companion the 
planet's orbit continues to evolve and the 
eccentricity can reduce from its peak. At the snapshot 
at 8\,Gyr at the end of the simulations, 
only $0.4-0.6\%$ of planets have an eccentricity 
above $0.84$ (Table~\ref{tab:1pl-kozai}, 
Figure~\ref{fig:1pl+binary}). That is, most 
planets which possess an eccentricity in the target 
range at some point do not do so at the end of the 
simulation: this is visualised in Figure~\ref{fig:sankey}
(middle panel).
We see from the 
orbital elements presented in Figure~\ref{fig:1pl+binary}
that it is only close and eccentric binary companions 
that can excite a high eccentricity within the system 
lifetime: the characteristic Kozai timescale is 
given by, e.g., \cite{Innanen+97}
\begin{equation}
t_\mathrm{Koz} = P_\mathrm{b} 
\left(\frac{a_\mathrm{B}}{a_\mathrm{b}}\right)^3
\frac{M_\mathrm{A}}{M_\mathrm{B}}
\left(1-e_\mathrm{B}^2\right)^{3/2},
\label{eq:tKoz}
\end{equation}
where $P_\mathrm{b}$ is the planet's orbital period. 
The locus where this estimated timescale is $8$\,Gyr
is shown in Figure~\ref{fig:1pl+binary}.

The simulations we ran with the additional weighting on 
binary semimajor axis from \cite{Raghavan+10} were more 
successful than those without the additional weighting. 
This is because, with a higher fraction of closer
binaries in these weighted runs, the distribution of Kozai timescales 
is shifted to shorter times.

\subsection{Combined scattering and Kozai}

\begin{table*}
  \caption{Outcomes for the runs with two planets
    with a binary companion. For each simulation set
    we show the number of runs $n_\mathrm{runs}$;
    the number where the planets underwent an 
    orbital instability $n_\mathrm{unstable}$;
    the number where the planet attained an
    eccentricity above $0.84$ at any point in the
    integration $n_{e\ge0.84}$(ever); the number
    where the planet had an eccentricity greater than
    $0.84$ at the end of the simulation
    $n_{e\ge0.84}$(8 Gyr); and the success fraction
    $f_\mathrm{success}=n_{e\ge0.84}\textrm{(ever)}/n_\mathrm{runs}$.}
  \label{tab:2pl-kozai}
  \begin{tabular}{lccccc}
    Simulation set & $n_\mathrm{runs}$ & $n_\mathrm{unstable}$ & 
    $n_{e\ge0.84}$ &
      $n_{e\ge0.84}$ & $f_\mathrm{success}$\\
    & & & (ever) & (8Gyr) & \\
      \hline
      2pl+binary-unweighted &  500 &  362 &   94 &  46 & 12.7\%\\
      2pl+binary-weighted   & 5000 & 3597 & 1082 & 521 & 14.5\%
  \end{tabular}
\end{table*}

\begin{figure}
  \includegraphics[width=0.50\textwidth]{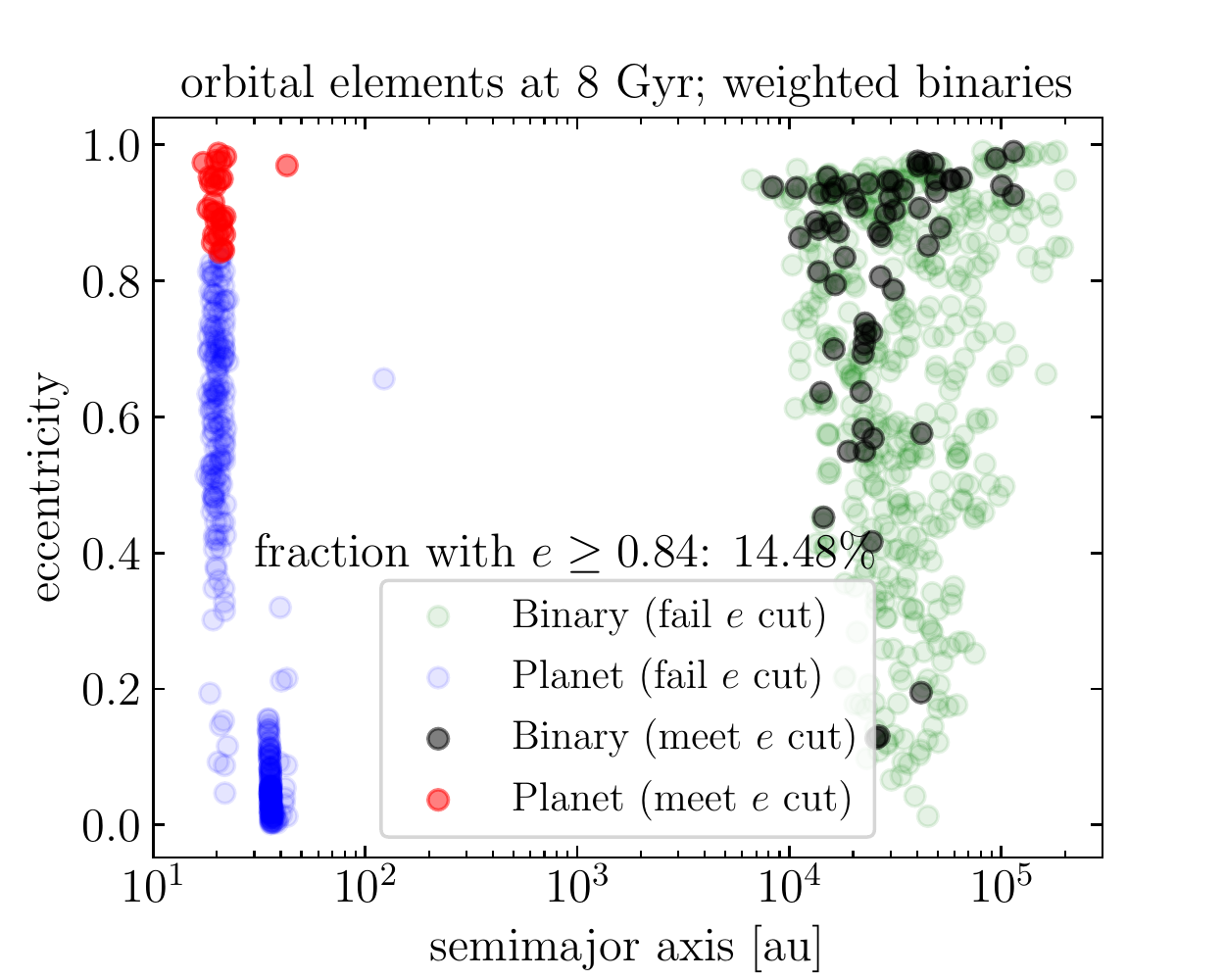}
  \caption{Orbital elements at 8\,Gyr for the
    systems with two planets , together with a binary
    companions drawn from the weighted posteriors of \protect\cite{Blunt+19}.
    500 out of 5\,000 simulations run are shown.}
  \label{fig:2pl+binary}
\end{figure}

We now study systems that initially comprise the primary star,
two planets, and the binary companion. In these systems, 
Lidov--Kozai cycles can act to further modulate the 
surviving planet's eccentricity after scattering; 
we see (Table~\ref{tab:2pl-kozai}) 
that this leads to significantly more planets 
having a high eccentricity at the end of the 
simulation than either scattering alone or binary 
perturbations alone.

An example of a successful system in this scenario
  is displayed in the right-hand panel of Figure~\ref{fig:egs}.
  Here, the two planets undergo an instability, removing
  one planet and leaving the survivor with an eccentricity
  a little below the target value. This survivor then
  undergoes Lidov--Kozai perturbations imposed by the
  binary, periodically raising it eccentricity above the target
  value. While we still require a favourable epoch of observation
  to catch the planet at a high eccentricty, the eccentricity
  is above the target value for a much larger fraction of
  the cycle than in the case of the single planet excited from
  low eccentricity.

The semimajor axis and eccentricity of surviving 
planets and stars 
at 8\,Gyr are shown in Figure~\ref{fig:2pl+binary}. 
$10.4\%$ of all systems have a planet with $e\ge0.84$,
and $14.5\%$ of the unstable runs --
an order of magnitude higher than 
the single-planet plus binary case, and also considerably 
higher than the equal-mass scattering with no binary companion. 
As in the case of the single planet plus binary, 
the binary companions in systems that reach $e\ge0.84$ are preferentially 
on more eccentric orbits with smaller semimajor 
axes\footnote{Two-sample Kolmogorov--Smirnov tests give $p$-values of $p=8\times10^{-15}$ 
  when comparing semimajor axis 
  distributions of binaries where the planet 
  does vs.\ does not end up at $e\ge0.84$, 
  and $p=5\times10^{-7}$ when comparing eccentricity distributions.}.

As seen in Figure~\ref{fig:sankey}
(bottom panel), 
now about half of the planets that ever have $e\ge0.84$ 
(following scattering) still have $e\ge0.84$ at 8\,Gyr, 
a much higher fraction than in the single-planet Kozai 
case. We can understand both this and the higher fraction 
of planets that attain high eccentricities by referring 
to the phase portrait for Kozai cycles, as discussed in 
Section~\ref{sec:discuss}.

\subsection{Relaxing the success criteria}

We then repeated the calculations of success rates for 
a target eccentricity of $e=0.8$, $1\sigma$ below the observed value. 
This modestly increased the success rates of all runs: 
the single-planet weighted binaries had a success rate of $0.9\%$; 
the two-planet weighted binaries $19.5\%$; and the two-planet 
pure scattering rates of $6.8\%$, $8.9\%$, $11.2\%$ and 
$5.9\%$ for mass ratios $\mu = 1$, $1.2$, $1.5$ and $2$ respectively.
We also relaxed the eccentricity target to $0.672$ (20\% below
  that observed), allowing us to account for the possibility of
  an underestimated formal error on the eccentricity
  measurement.
  Here, we found success rates of $1.8\%$ for single-planet
  weighted binaries, $41.7\%$ for two-planet weighted binaries, and
  $33.2\%$, $31.8\%$, $23.1\%$ and $9.2\%$ for two-planet scattering. On the whole,
the combination of scattering and Lidov--Kozai perturbations remains 
the most effective.

Finally, we discuss relaxing the time constraint. While we
  consider the most useful definition of a success to be that the
  eccentricity is suitably high at a given epoch of observation,
  it is also worth exploring the fraction of systems that ever
  attain such a high eccentricity. For the single-planet weighted
  binary simulations, we find that 317 out of 5000 systems ever attain
  $e\ge0.84$ (about 10 times the number at 8\,Gyr), while for the
  two-planet weighted binary simulations, we find 1082 systems
  attain $e\ge0.84$ (only about twice the number at 8\,Gyr).
  The combination of scattering and Lidov--Kozai perturbations
  therefore increases the fraction of systems ataining high
  eccentricity at any time, but it also significantly increases
  the ``duty cycle'' at which these systems have high eccentricity,
  making them more likely to be observed in the high-eccentricity state.

\section{Discussion}

\label{sec:discuss}

\subsection{Why Kozai plus scattering works well}

\begin{figure*}
  \includegraphics[width=0.33\textwidth]{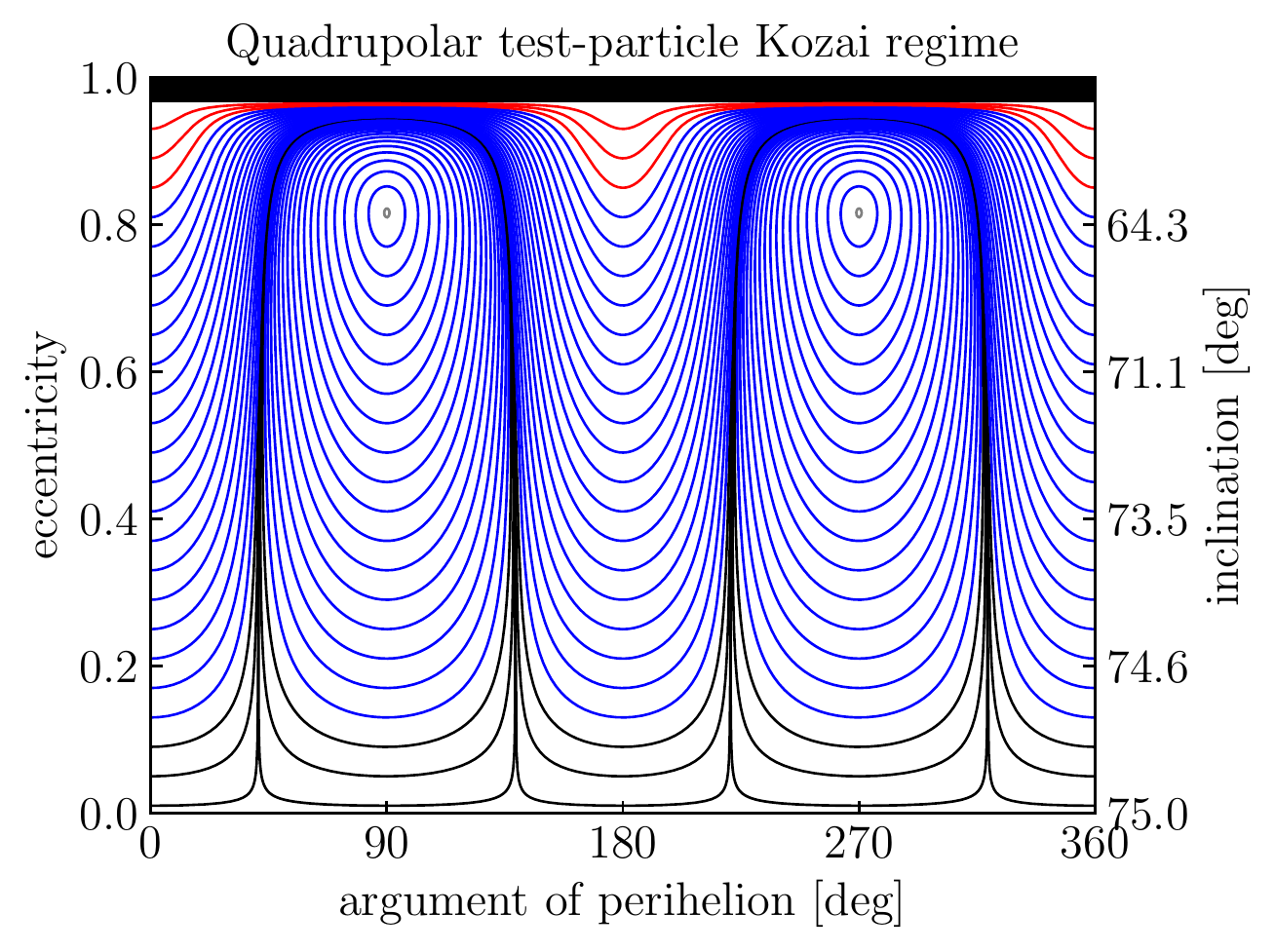}
  \includegraphics[width=0.33\textwidth]{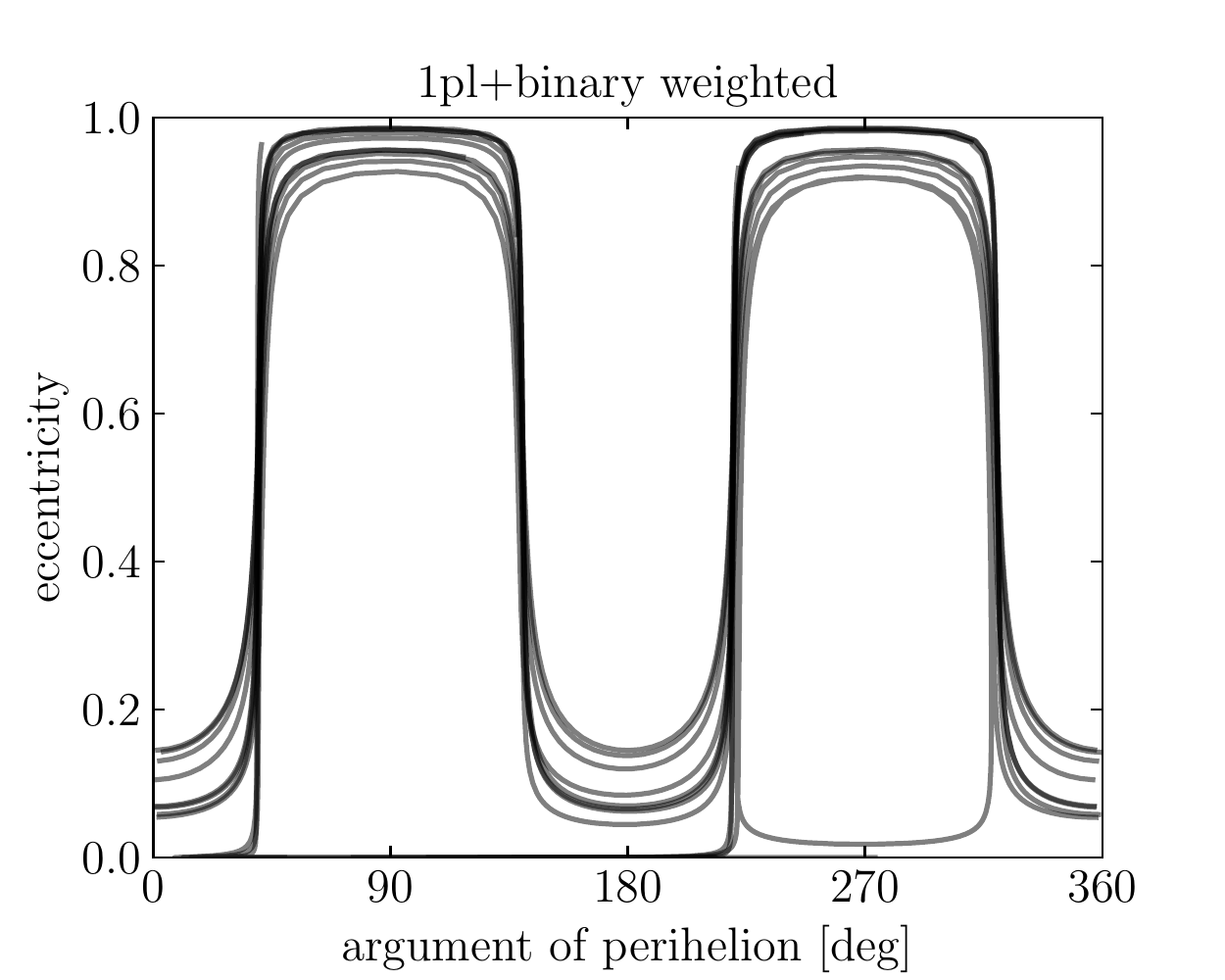}
  \includegraphics[width=0.33\textwidth]{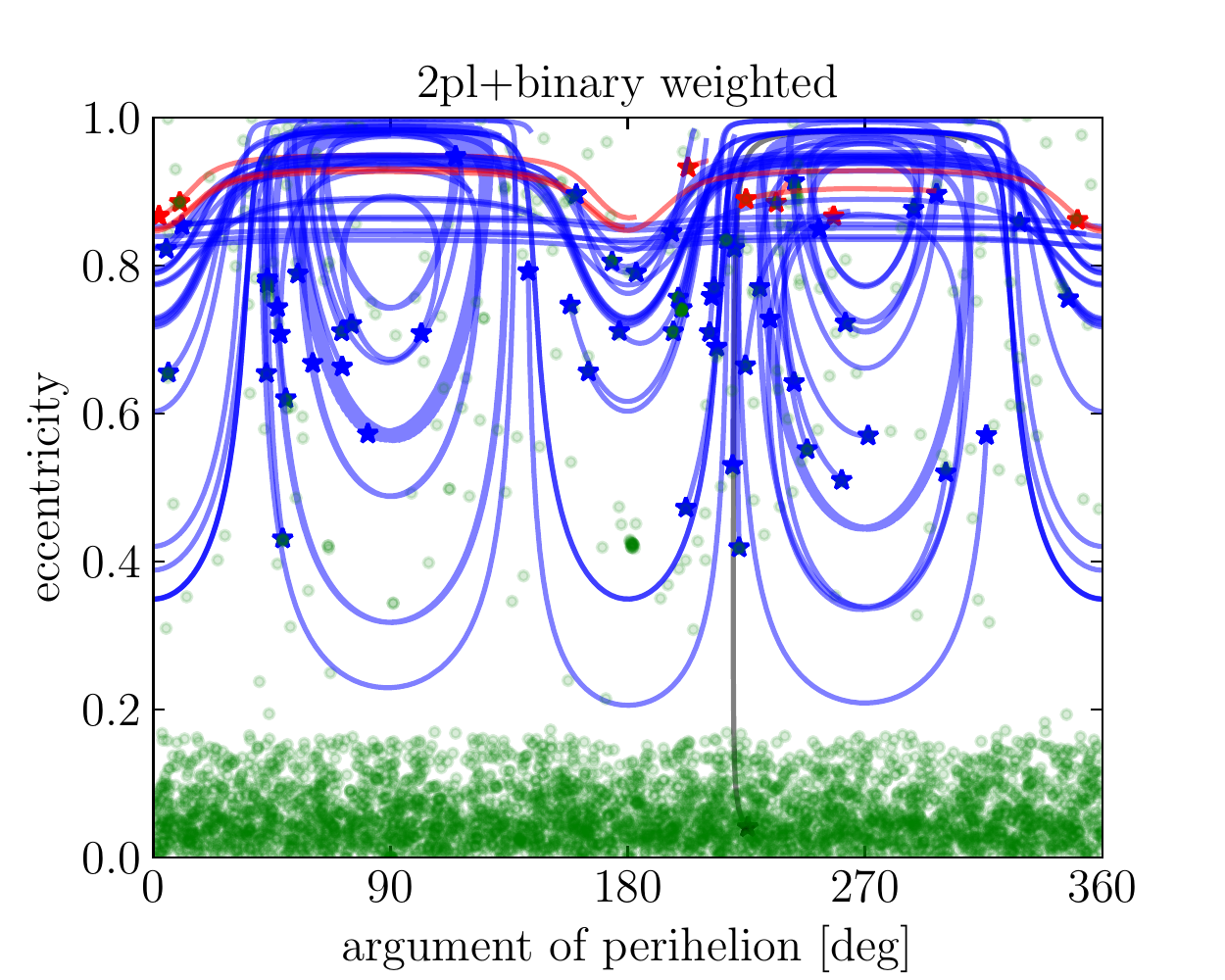}
  \caption{Left: phase portrait for the classical Lidov--Kozai 
    resonance in the quadrupolar approximation with 
    an interior test particle, with a constant 
    $C_\mathrm{Koz}=\sqrt{1-e^2}\cos I=0.258806$. Eccentricity 
    is plotted against the argument of pericentre 
    ($\omega$, a measure of orbital orientation), 
    and orbits evolve along the curves. 
    The right-hand ordinate axis shows the value of 
    inclination corresponding to the eccentricity 
    on the left-hand axis.
    Black trajectories 
    begin below $e=0.1$ and at some point attain $e\ge0.84$;
    Blue trajectories begin between $e=0.1$ and $e=0.84$ 
    and at some point attain $e\ge0.84$; red trajectories 
    are always above $e=0.84$; and grey trajectories 
    (librating round the fixed points at $\omega=90^\circ$
    and $270^\circ$) never attain $e\ge0.84$.
    The black region at $e\approx1$ is forbidden by this 
    value of $C_\mathrm{Koz}$. 
    Centre and right: Planetary orbital trajectories in the space of argument of 
    pericentre $\omega$
    and orbital eccentricity $e$. We show systems where the surviving 
    planet had $e\ge0.84$ at the end of the simulation, from the 
    simulation sets 1pl+binary-weighted (centre) and 
    2pl+binary-weighted (right). In the right-hand panel, 
    orbital elements are shown in green while two planets are still 
    present in the system, and the orbital evolution is dominated 
    by planet--planet secular or scattering interactions (Here the
    evolution is typically more rapid than our time sampling of the
    simulations, and we therefore mark the orbital elements as disconnected
    points). Orbital 
    elements are shown in red, blue or black after the loss of a 
    planet, with the same scheme as in the left panel; a star marks the 
    first output time after a planet is lost and the orbit evolves solely 
    under the influence of the binary companion. Note how in all but one case 
    is the eccentricity at the start of this stage above $0.4$. In 
    the centre and right-hand panels, 500 out of the 5\,000 simulations run are 
    sampled.}
  \label{fig:phase}
\end{figure*}

Here we discuss why the combination of 
scattering and Kozai works much better than Kozai alone 
starting from a single low-eccentricity planet. 
The peak planetary eccentricity attainable in a Kozai cycle 
depends on the binary inclination and the initial planetary 
eccentricity. A higher initial planetary eccentricity, or a higher 
binary inclination, usually means 
a higher peak eccentricity for the planet.

We illustrate this in Figure~\ref{fig:phase}. 
In the left-hand panel we show the phase portrait 
for the classical quadrupolar Lidov--Kozai 
scenario of a massless interior particle and a massive 
outer binary. We choose a Kozai constant 
$C_\mathrm{Koz}=\sqrt{1-e^2}\cos I=0.258806$, corresponding 
to an orbital inclination of $75^\circ$ at $e=0$. The phase 
portrait shows the evolution of orbits in the phase 
space of argument of perihelion and eccentricity. Trajectories 
are colour-coded depending on their initial and 
peak eccentricities.
Black trajectories are those that start at low eccentricity 
($e<0.1$) and later reach $e=0.84$; this is the region of 
phase space where an inclined binary is effective at exciting 
a single planet to the required eccentricity. Scattering among 
comparable-mass planets excites higher $e$, and here there are three 
regimes. Blue trajectories spend part of their time at $e\ge0.84$, 
and part at $e<0.84$; this shows a region of phase space where 
planet--planet scattering alone does not produce the 
required eccentricity, but later Kozai perturbations do.
Red trajectories spend all of their time at $e\ge0.84$: 
if planet--planet scattering lands a planet in this region of 
phase space, it will always be observed at the 
requisite eccentricity. 
Grey trajectories (seen enclosing the fixed points 
at $\omega=90^\circ$ and $\omega=270^\circ$) 
never attain $e\ge0.84$; 
planet--planet scattering could land a planet in the regions 
close to these fixed points, 
but the region is comparatively small. 
Overall, we see that planet--planet scattering gives 
an initial eccentricity boost to the surviving planet, 
which makes it easier to attain a still higher 
eccentricity later on under the influence of Lidov--Kozai
perturbations.

In the phase portrait of Figure~\ref{fig:phase}, 
eccentricity and inclination are complementary, 
as shown by the secondary ordinate axis: 
a higher eccentricity means a lower inclination. 
Thus, to attain the requisite $e$ from a higher initial 
eccentricity requires a lower inclination with respect to 
the binary. This means that there is less restriction 
on the binary orbital inclination when the initial 
eccentricity is higher, increasing 
the likelihood that a planet after scattering 
will be forced up to the observed eccentricity.

In the center and right panels of Figure~\ref{fig:phase}
we show the equivalent ``phase space'' from our $N$-body
runs\footnote{We note that these are not true phase portraits, as the
energy is different for each trajectory.}.
We show the orbital trajectories for planets that have
$e\ge0.84$ at the end of the simulations, for simulation sets
1pl+binary-weighted (centre) and 2pl+binary-weighted 
(right). In the right-hand 
panel, the first data output following the loss of a 
planet (after which the remaining planet evolves 
purely under Lidov--Kozai forcing from the binary) 
is shown as a star.
In the single-planet simulations, all 
planets start at low eccentricity, and only a small 
number have a suitable binary inclination that allows them 
to reach $e=0.84$. In the two-planet simulations, after one 
planet is lost and the scattering has ended, the surviving 
planet often already has a moderately large eccentricity. 
This allows a larger peak eccentricity to be attained. The 
colour-coding in the right-hand panel is as in the left-hand panel. 
Note in particular that only one planet (of the 500 systems sampled) is excited to 
high eccentricity from $e<0.1$: all of the others end the 
phase of scattering with an eccentricity of at least $0.4$.
  
\subsection{Mutual inclination in the 1pl+binary and 2pl+binary cases}

\begin{figure}
  \includegraphics[width=0.5\textwidth]{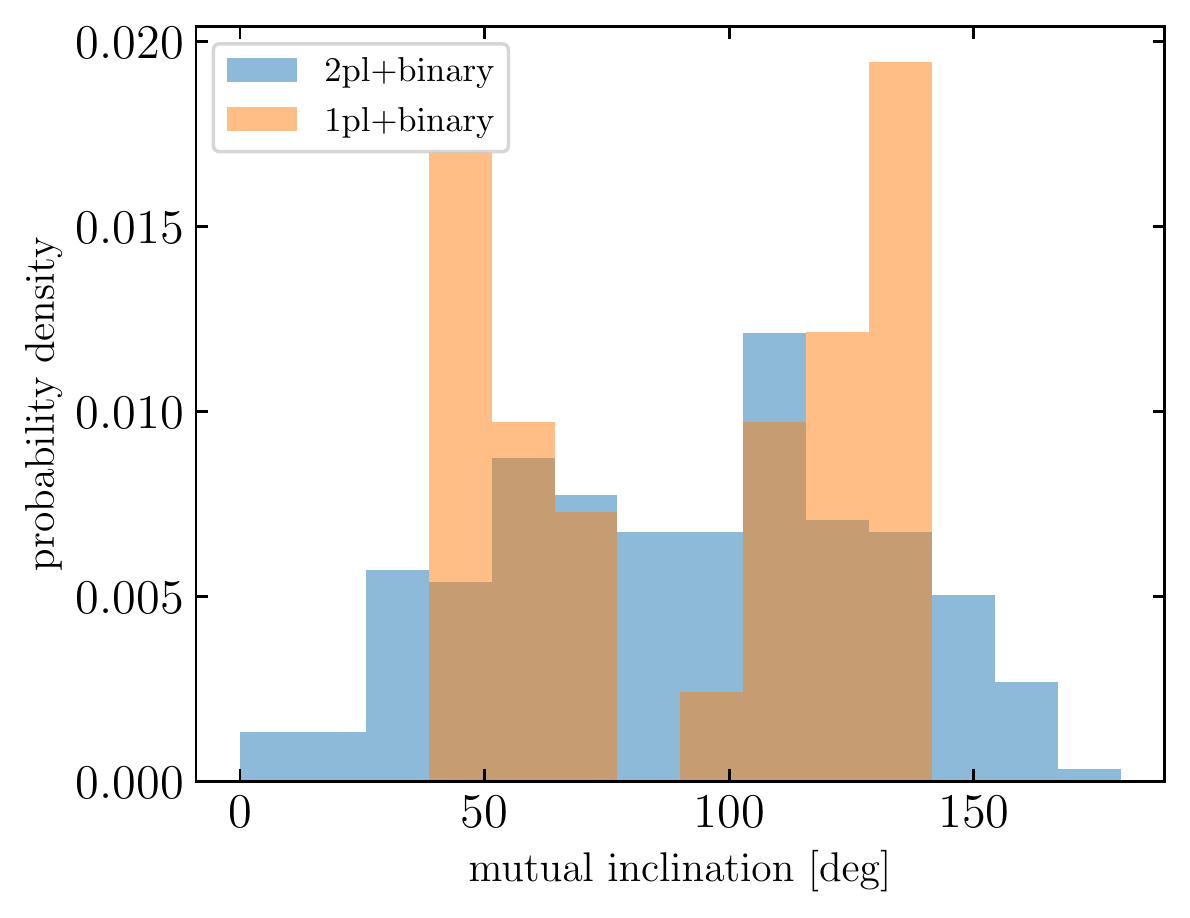}
  \caption{Mutual inclination between the orbit of the binary 
    and of the planet, for systems where the latter has 
    $e\ge0.84$ at the end of the simulation.}
  \label{fig:imut}
\end{figure}

One potential observational means of distinguishing between 
the pure Kozai, versus the scattering plus Kozai, dynamical 
histories, lies in the mutual inclination of the planet 
and the binary companion at the present time. The probability 
distributions of mutual inclination at the end of the 
simulation are shown in Figure~\ref{fig:imut}, for 
the 1pl+binary-weighted and the 2pl+binary-weighted 
simulation sets where the surviving planet ends with
an eccentricity $\ge0.84$. We see that in the single-planet 
case, where the orbital evolution is heavily constrained by starting 
at low-eccentricity, the range of mutual inclinations 
in the high-eccentricity systems is rather limited, with 
both polar and low-inclination configurations being 
prohibited. In contrast, in the two-planet systems, the 
final mutual inclination between the surviving eccentric 
planet and the binary roughly follows an isotropic distribution. 
Thus, if the inclination of the planet's orbit can be 
measured or constrained (by astrometric measurement of 
the star's reflex motion with \emph{Gaia,} or else 
by direct imaging), a polar or low-inclination 
configuration at the present day would be inconsistent with the 
single-planet origin, and support an earlier scattering 
history.

\subsection{Effects of short-range forces}

We have modelled the orbital dynamics with 
  pure Newtonian point-mass dynamics. However, in 
highly-eccentric systems short-range 
forces can become significant. Two such forces are 
those that arise from general relativity, and 
from tidal deformation of the bodies.

Orbits in the relativistic two-body problem 
are subjected to a slow precession due to the leading-order 
effects of general relativity. In the three-body problem, 
this can suppress the excitation of eccentricity 
through Lidov--Kozai cycles \citep[e.g.,][]{Holman+97}. 
The precession rate 
of an eccentric orbit under GR precession is given by 
\citep[e.g.,][]{Naoz16}:
\begin{equation}
\frac{\mathrm{d}\omega}{\mathrm{d}t}=
\frac{3k^3M_\star^{3/2}}{a^{5/2}c^2\left(1-e^2\right)},
\end{equation}
where $k$ is the Gaussian gravitational constant 
and $c$ the speed of light. This precession rate yields 
a timescale for one cycle of precession of 
42\,Gyr, for the planet at 18\,au, if its eccentricity 
is zero, and 12\,Gyr if its eccentricity is $0.84$. 
As the timescale for Lidov--Kozai cycles must be 
$\lesssim8$\,Gyr, the system age, to significantly 
change the planet's eccentricity, the effects of GR 
should not significantly affect the Lidov--Kozai dynamics
of those systems that are affected on astrophysically 
relevant timescales.

Tidal distortion of one or both bodies can also
  cause apsidal precession. The rate is given by 
\begin{eqnarray}
  \frac{\mathrm{d}\omega}{\mathrm{d}t}&=&
  \frac{15\sqrt{\mathscr{G}\left(M_\star+M_\mathrm{b}\right)}}{8a^{13/2}}
  \frac{8+12e^2+e^4}{\left(1-e^2\right)^5}\\
  &&\left(\frac{M_\mathrm{b}}{M_\star}k_\star R_\star^5
  +\frac{M_\star}{M_\mathrm{b}}k_\mathrm{b} R_\mathrm{b}^5\right),
\end{eqnarray}
where $k_\star$ and $k_\mathrm{b}$ are the Love numbers of the
star and the planet, and $\mathscr{G}$ is the gravitational
constant \citep{FabryckyTremaine07}. At $e=0.84$, we find
for this system a precession time of $1.5\times10^{17}$\,yr for one cycle,
when assuming a planetary radius of 1 Jupiter radius and
unrealistically high values of $k_\star = k_\mathrm{b} = 1$.
Hence, for our purposes, tidal effects are negligible,
although they would affect the small number of planets
that end up colliding with the star, and the rate of
formation of Hot Jupiters through the different
dynamical channels.

\subsection{Caveats on orbital evolution of the binary}

In this paper, we have treated the orbit of the binary as unchanging.
This has enabled us to isolate the effect Lidov--Kozai
cycles from a binary perturber on a fixed orbit have 
on the planet after scattering, without worrying 
about changes to the binary orbit through the Galactic 
tide or stellar flybys, or indeed about the breakup of 
the binary by passing stars. Here, we note that 
the binary has clearly survived several Gyr without 
disruption. Binaries at $a=10^4$\,au are disrupted at a rate of
$\sim50\%$ per $10\mathrm{\,Gyr}$ \citep{Kaib+13,Correa-OttoGil-Hutton17}. 
While the binary has avoided this, it may have had its 
orbit changed, a process more likely if the stars were formed 
at smaller Galactocentric radii than their present orbit 
where field star densities are higher, before migrating 
outwards \citep{Wielen+96,SellwoodBinney02,
  NievaPrzybilla12,Kubryk+15,Frankel+18}. Formation 
closer to the Galactic centre may be likely based on 
the star's slightly super-Solar metallicity, but the extent 
(and even necessity) of migration for the Sun at least is 
disputed \citep[e.g.,][]{Minchev+18,Haywood+19,Frankel+20}, 
depending on assumptions about the enrichment 
history of the ISM. \cite{Kaib+13} show that a doubling or quadrupling 
of field star densities, as would be expected if formation were
several kpc closer to the Galactic centre, leads to a moderate increase 
in the fraction of planetary systems destabilised by wide binaries; 
however, \cite{Correa-OttoGil-Hutton17} show that, 
despite perturbations to their
orbits, even highly eccentric binaries at 
$\sim10^4$\,au are unlikely to directly destabilise planets
within 30\,au. Hence, the binaries most effective at exciting
high eccentricity through the Kozai mechanism
(having lower $a$ and higher $e$), and which have
avoided being broken up by flybys in the field,
will also have avoided penetrating too deeply into the
inner system to directly destabilise any planets. 
The most likely effect of considering changes to 
the binary orbit, then, would be some modulation of 
the Lidov--Kozai timescale as the binary orbital elements 
are perturbed.

\subsection{Other systems}

The mechanism of eccentricity excitation by 
scattering plus Lidov--Kozai interaction 
that we have described in this paper will not be limited 
to the HR~5183 system. While close (within a few tens of au)
binary companions are expected to hinder the planet 
formation process \citep[e.g.,][]{Jang-Condell15,MarzariThebault19}, this is not 
expected for the wider binary systems. Indeed, 
observationally, the binary companions of exoplanet 
systems seem to have a wider semimajor axis distribution 
than the companions of field stars chosen without 
reference to their exoplanet systems. This is seen in 
studies of transiting planetary systems 
\citep{Kraus+16,Ziegler+20,Howell+21}, 
as well as of giant planets within a few au 
of their host star \citep{Fontanive+19,Hirsch+20}. 
Meanwhile, the masses and semimajor axes of planets in 
binaries wider than 1\,000\,au are the same as those of planets
orbiting single stars \citep{Fontanive21}. 
Of particular relevance is the study of the binary 
companions to white dwarfs with photospheric metal lines by \cite{Zuckerman14}: 
these ``polluted'' white dwarfs are thought to be displaying 
the traces of asteroids or planetary remnants scattered into the 
white dwarf's atmosphere 
\citep[e.g.,][]{Alcock+86,DebesSigurdsson02,JuraYoung14,Farihi16}. 
The source 
region for these bodies must be several to several tens 
of au in order to survive the host star's asymptotic giant 
branch evolution \citep{MustillVillaver12,Mustill+18} -- exactly the semimajor axis 
range of HR~5183b. For these systems, \cite{Zuckerman14} found 
that binary companions beyond 1000\,au seem 
not to suppress the existance of planetary systems; a similar result 
was found by \cite{Wilson+19}. This raises 
hopes that more systems similar to HR~5183, with a planet at ~10\,au 
and a wide binary beyond 1000\,au, may be found. 

Indeed, the NASA Exoplanet Archive\footnote{\url{https://exoplanetarchive.ipac.caltech.edu/},\newline 
accessed 2021-01-28.} lists 16 planets with an eccentricity $e\ge0.8$. Of these, 
8 are listed as being in multiple star systems (HD~28254~b, \citealt{Naef+10}; 
HR~5183~b, \citealt{Blunt+19}; HD~108341~b, \citealt{Moutou+15}; 
HD~156846~b, \citealt{Stassun+17}; HD~4113~b, \citealt{Tamuz+08}; 
HD~7449~b, \citealt{Wittenmyer+19}; HD~80606~b, \citealt{Stassun+17}; and 
HD~20782~b, \citealt{Udry+19})\footnote{Queries of \emph{Gaia} EDR3 
\citep{Gaia+20} 
did not reveal any comoving companions to the other eight stars.}. While this 
is a small and possibly biased sample, this fraction of high-eccentricity
planets with binary companions (50\%) is much higher than the fraction of 
all planetary systems in multiple stellar systems (321 out of 3213, or 10\%). 
This suggests a role of stellar binarity in generating these very large 
eccentricities, which we have shown is easier if the planetary system 
began with multiple planets and underwent scattering.

Finally, we note that \cite{KaneBlunt19} showed that it
  is possible for a terrestrial planet to remain on a stable
  orbit in the Habitable Zone of the HR5183 system, under the
  perturbations from the known planet on its current orbit.
  However, \cite{Matsumura+13,Carrera+16,Kokaia+20} have shown
  that treating a known planet's orbit as fixed gives an optimistic
  estimate of the survivability of Habitable Zone planets, in
  the case of planet--planet scattering. Future studies should
  re-evaluate the prospects for the existence of such planets
  in HR5183 and similar systems under the different possible
  dynamical evolution scenarios.

\section{Conclusions}

\label{sec:conclude}

We have examined and compared the efficiencies of three mechanisms 
of exciting an exoplanet's orbit to high eccentricity:
scattering in an unstable two-planet system, Lidov--Kozai 
forcing from a wide stellar binary companion, and a combination 
of planet--planet scattering plus Lidov--Kozai cycles 
from a wide binary. We have reproduced the 
current orbital configuration of the HR~5183 system with these 
three mechanisms, with the following quantitive success rates 
(planetary eccentricity at the end of the simulation $\ge0.84$, its current value):
\begin{itemize}
\item Two-planet scattering: $2.8-7.2\%$ of unstable 
  systems end up with a planet 
  with $e\ge0.84$. The success rate depends on the mass ratio between the 
  outer and inner planets, being lowest for equal-mass planets and
  peaking for a mass ratio $\sim1.5$.
\item Single-planet Lidov--Kozai: $0.6\%$ of systems end up with a planet 
  with $e\ge0.84$.
\item Two-planet scattering plus Lidov--Kozai: 
  $14.5\%$ of unstable systems
  with equal-mass planets end up with a planet with $e\ge0.84$.
\end{itemize}

Given the existence of the comoving and probably bound companion
star to HR5183, we consider that planet--planet scattering 
followed by Lidov--Kozai cycles driven by the binary 
is the best explanation of the planet's high eccentricity. 
This can potentially be tested if the inclination of the planet's 
orbit can be measured, since eccentricity excitation by Lidov--Kozai 
cycles from a circular orbit in a single-planet system leads to 
neither coplanar nor polar orbits, whereas Lidov--Kozai cycles 
acting after planet--planet scattering has occurred permit 
any mutual inclination between the surviving planet and the 
binary companion. Given that wide binaries such as the 
system we have studied are observed to 
happily coexist with planetary systems, this combination of 
planet--planet scattering and Lidov--Kozai cycles may 
affect the eccentricities of many wide-orbit 
giant planets. Indeed, the existing small sample of highly-eccentric 
planets suggests that these are more likely to be found in binary stellar 
systems.

\section*{Acknowledgements}

The authors thank Stephen Kane, Antoine Petit, and the anonymous 
  referee for comments that improved the manuscript.
AJM and MBD acknowledge support from project grant 2014.0017 ``IMPACT'' 
from the Knut and Alice Wallenberg Foundation. AJM 
acknowledges support from Career grant 120/19C from the 
Swedish National Space Agency. 
The simulations were performed on resources provided by the Swedish 
National Infrastructure for Computing (SNIC) at Lunarc, 
partially funded by the Swedish Research Council 
through grant agreement no. 2016-07213. 
This research made use of Astropy,\footnote{\url{http://www.astropy.org}} 
a community-developed core Python package for 
Astronomy \citep{astropy:2013, astropy:2018}. 
This research made use of NumPy \citep{2020NumPy-Array}, 
SciPy \citep{2020SciPy-NMeth} and MatPlotLib \citep{2007CSE.....9...90H}.
This work has made use of data from the European Space Agency (ESA) mission \emph{Gaia}
(\url{https://www.cosmos.esa.int/gaia}), processed by the \emph{Gaia} Data Processing and Analysis Consortium
(DPAC, \url{https://www.cosmos.esa.int/web/gaia/dpac/consortium}). Funding for the DPAC has been provided by national institutions, in particular the institutions participating in the \emph{Gaia} Multilateral Agreement. This research has made use of the NASA Exoplanet Archive, which is operated by the California Institute of Technology, under contract with the National Aeronautics and Space Administration under the Exoplanet Exploration Program.

\section*{Data availability statement}

The data underlying this paper will be made available 
on reasonable request to the corresponding author.

\section*{Note added in proof}

Following the acceptance of this paper, \cite{Venner+21} published an improved orbital fit for both the planet and the binary companion to HR5183, with tentative but not statistically significant evidence for orbital misalignment. Future astrometry should refine the orbital misalignment angle further.

\bibliographystyle{mnras}
\bibliography{HR5183_formation}

\bsp    
\label{lastpage}

\end{document}